\documentstyle[12pt,epsf]{article}
\newcommand{\be}{\begin{eqnarray}}
\newcommand{\ee}{\end{eqnarray}}

\catcode`\@=11
\def\lsim{\mathrel{\mathpalette\@versim<}}
\def\gsim{\mathrel{\mathpalette\@versim>}}
\def\@versim#1#2{\vcenter{\offinterlineskip
\ialign{$\m@th#1\hfil##\hfil$\crcr#2\crcr\sim\crcr } }}
\catcode`\@=12

\begin{document}
\pagestyle{empty}    
 
\begin{flushright}
KANAZAWA-02-, IFUNAM-FT2002-05
 \\ November 2002
 \end{flushright}

 \begin{center}
{\Large\bf Suppressing FCNC and CP-Violating Phases\\
with Extra Dimensions} \footnote{
Presented by J.K. at SUSY 2002, Hamburg.}
\end{center} 

\vspace{2cm}
\begin{center} {\sc Jisuke Kubo$^{(a,b)}$} 
    and {\sc Haruhiko  Terao$^{(b)}$} 
\end{center}

\begin{center}
{\em (a) Instituto de F\' isica,  UNAM,
Apdo. Postal 20-364,
M\' exico 01000 D.F., M\' exico\\
(b) Institute for Theoretical Physics, 
Kanazawa  University, \\
Kanazawa 920-1192, Japan
}
\end{center}

\vspace{3cm}
\begin{center}
{\sc\large Abstract}
\end{center}

\noindent
The infrared attractiveness of 
soft-supersymmetry breaking (SSB) parameters
is investigated in extra dimensions.
The SSB parameters are assumed to run from
the Planck scale $M_{\rm PL}$ down to the GUT scale
$M_{\rm GUT}$.
We find that, even for $M_{\rm PL}/M_{\rm GUT}
=100$, the
SSB parameters  in six dimensions can align 
out of their anarchical disorder at 
$M_{\rm PL}$ in such a way that flavor-changing neutral currents as well
as dangerous CP-violating phases are sufficiently suppressed at $M_{\rm GUT}$.
A concrete, realistic 
example based on $SU(5)$ in six dimensions is
presented.

\newpage
\pagestyle{plain}
It is widely accepted  that the effects of supersymmetry breaking appear
as soft supersymmetry breaking (SSB) terms.
However, renormalizability allows to
introduce more than 100 new parameters into the 
minimal supersymmetric standard model (MSSM).
The problem is not only this large number of
the independent parameters, 
but also the fact that 
the SSB terms induce
large flavor-changing neutral current (FCNC) processes and
CP-violating phases, which are
severely constrained by precision
experiments  \cite{fcnc-mueg,fcnc-k,fcnc-edm,fcnc-bsg,fcnc}.
Therefore, the huge degrees of
freedom involved in
the soft-supersymmetry breaking (SSB) parameters
have to be highly constrained in all viable
supersymmetric models.
This has been called the supersymmetric flavor problem.
To overcome  this problem, several ideas
of SUSY breaking and its mediation mechanisms
have been proposed; gauge mediation \cite{gauge},
anomaly mediation \cite{anomaly}, gaugino mediation \cite{gaugino}
and so on.
The common feature behind these ideas is that the leading parts of
the SSB parameters are given by flavor-blind radiative corrections.
It is noted that the anomaly mediation and the gaugino mediation work
on the assumption that the tree-level contributions for the 
SSB parameters at a fundamental scale $M_{\rm PL}$ are sufficiently
suppressed, {\it e.g.}, by sequestering of branes for the visible sector and 
the hidden SSB sector, since there is no reason for these terms to 
be flavor universal.
However, it has been  argued recently \cite{string}
that such a sequestering mechanism cannot be simply
realized in generic supergravity or superstring inspired models.
An interesting way out from this problem is to suppress the tree-level
contributions by certain field theoretical dynamics.
There have been indeed  several attempts along the line of
thought,  in which use has been made \cite{ns,nkt,ls}
that the SSB parameters
are suppressed in the infrared limit in
approximate superconformal field theories \cite{karch}.

In \cite{kubo}, we proposed another possibility
in more than four dimensions that
flavor-blind radiative corrections
are much more dominant than any other flavor non-universal contributions.
At this meeting we would like to present our idea and results.
The mechanism that we propose 
implements the power-law running of 
couplings \cite{veneziano,dienes1} in supersymmetric
field theories with $\delta$
extra compactified dimensions and
at the same time  the infrared attractiveness
of the SSB parameters \cite{ross}.
We consider the simplest case in which only
the non-abelian gauge supermultiplet
propagates in the $(4+\delta)$-dimensional bulk and the supermultiplets
containing the matter and Higgs fields are localized
at our 3-brane \cite{dienes1,antoniadis1,arkani1}.
In this mechanism  the gaugino mass $M$,
which is assumed to be generated at the fundamental scale $M_{\rm PL}$
by some SUSY breaking mechanism,
receives
a correction proportional to
$(M_{\rm PL}/M_{\rm GUT})^\delta$
at the grand unification scale $M_{\rm GUT}$,
and more importantly induces dominant flavor-blind corrections to other
SSB parameters.
The most interesting finding is that the squared soft-scalar masses
$(m^2)^i_j$
and the soft-trilinear couplings $h^{ijk}$
become  so aligned at $M_{\rm GUT}$
that FCNC processes and
dangerous CP-violating phases are sufficiently suppressed.
In this class of models,
all the A-parameters $h$'s,
B-parameter $B_H$ and  soft-scalar masses $m^2$'s
in the minimal supersymmetric standard model (MSSM)
are basically  fixed  as functions
of  the unified gaugino mass $M$ and the $\mu$-parameter
$\mu_H$,  up to corrections coming from Yukawa interactions.
Therefore, this class of models can
not only  overcome
the supersymmetric flavor problem, but also
have a  large predictive power.
Moreover, 
no charged sparticles become tachyonic
in these models.

Let us be more specific.
As mentioned, we assume that the $(4+\delta)$-dimensional
gauge supermultiplet
propagates in the bulk, and  all the
$N=1$ chiral supermultiplets $\Phi_i =(\phi_i,~\psi_i)$
containing  matters and
Higgses propagate only in four dimensions.
The gauge supermultiplet contains
a chiral supermultiplet $ \Gamma $ in the adjoint
representation, where we assume that $\delta$ is equal to one or two.
We assign an odd parity to $\Gamma$
so that it does not contain zero modes \cite{dienes1,arkani1},
and do not have any interactions
with $\Phi$'s.
To simplify the situation we further assume that
each extra dimension is compactified on
a circle with the same radius $R$.
The size of $R$ is model dependent, but
throughout this paper we
assume that $M_{\rm GUT}=1/R$.
With these assumptions,  the boundary superpotential
has a generic form
\begin{equation}
W(\Phi) = \frac{1}{6} Y^{ijk} \Phi_i \Phi_j \Phi_k +
\frac{1}{2} \mu^{ij} \Phi_i\Phi_j ,
\end{equation}
and that the SSB Lagrangian $L_{\rm SSB}$ can be written as
\begin{equation}
-L_{\rm SSB}= \left(\frac{1}{6}
 h^{ijk} \phi_i \phi_j \phi_k
 +  \frac{1}{2}  B^{ij} \phi_i \phi_j
+  \frac{1}{2} \sum_{n=0} M\lambda_n \lambda_n+
\mbox{h.c.}\right)
+\phi^{*j}(m^2)^i_j \phi_i,
\label{ssbL}
\end{equation}
where $\lambda_n$'s are the Kaluza-Klein modes
of the gaugino, and
we have assumed  a unique gaugino mass $M$
for all $\lambda$'s.

We  consider the renormalization group (RG) running of the parameters
between the fundamental scale
$M_{\rm PL}=M_{\rm Planck}/\sqrt{8\pi}\simeq
2.4 \times 10^{18}$ GeV and
$M_{\rm GUT}$.
To see the gross behavior of
the RG running, we first consider
the contributions coming
from only the gauge supermultiplet,
because it is  the only
source responsible for the power-law running \cite{veneziano,dienes1}
of the parameters under the assumptions specified above.
In the flavor bases in which couplings of the gauginos are diagonal,
only diagonal elements of the anomalous dimensions can contribute.
We find the following set of the  one-loop  $\beta$ functions
in this approximation \cite{dienes1,kobayashi1}:
\begin{eqnarray} 
\Lambda \frac{d g }{d\Lambda} &=&
=-\frac{2}{16\pi^2}C(G)G_\delta^2g~,~
\Lambda \frac{d M }{d\Lambda} =
-\frac{4}{16\pi^2}  C(G)G_\delta^2 M,
\label{bg1}\\
\Lambda \frac{d Y^{ijk} }{d\Lambda}&=&
-\frac{2}{16\pi^2} (C(i)+C(j)+C(k))G_\delta^2 Y^{ijk},
\label{bY1}\\
\Lambda \frac{d \mu^{ij} }{d\Lambda} &=&
-\frac{2}{16\pi^2} (C(i)+C(j))G_\delta^2~\mu^{ij},
\label{bmu1}\\
\Lambda \frac{d B^{ij} }{d\Lambda} &=&
\frac{2}{16\pi^2}(C(i)+C(j))G_\delta^2
(2 M\mu^{ij}-B^{ij}),
\label{bb1}\\
\Lambda \frac{d h^{ijk} }{d\Lambda} &=&
\frac{2}{16\pi^2} (C(i)+C(j)+C(k))G_\delta^2 (2 M Y^{ijk}-h^{ijk}),
\label{bh1}\\
\Lambda \frac{ d (m^2)^i_j }{ d\Lambda } &=&
- \frac{8}{16\pi^2} C(i)\delta_j^i G_\delta^2|M|^2,
\label{bm21}
\end{eqnarray}
where
$G_\delta = g X_\delta^{1/2}(R\Lambda)^{\delta/2}$, and 
$
X_\delta =\pi^{\delta/2} \Gamma^{-1}(1+\delta/2)
= 2(\pi)~~\mbox{for} ~~\delta=1(2)$ \cite{dienes1}
\footnote{$X_\delta$ is regularization scheme
dependent. See \cite{kubo1} for a detailed
analysis on the regularization dependence.}

The gauge coupling is denoted by $g$, and
$C(G)$ stands for the
quadratic Casimir of the adjoint representation of the
gauge group $G$, and $C(i)$
for  that of the representation $R_{i}$.
It is easy to show that the evolution of $Y^{ijk}, \mu^{ij}$
and $M$ are related to that of $g$ as
\begin{eqnarray}
M(M_{\rm GUT}) &=& 
\left(\frac{g(M_{\rm GUT})}{g(M_{\rm PL})}\right)^{2}
M(M_{\rm PL})~,~
Y^{ijk}(M_{\rm GUT}) =
\left(\frac{g(M_{\rm GUT})}{g(M_{\rm PL})}\right)^{\eta_Y^{ijk}}
Y^{ijk}(M_{\rm PL}),
\label{yijk}\\
\mu^{ij}(M_{\rm GUT}) &=&
\left(\frac{g(M_{\rm GUT})}{g(M_{\rm PL})}\right)^{\eta_\mu^{ij}}
\mu^{ij}(M_{\rm PL}),
\label{muij}
\end{eqnarray}
where
\begin{eqnarray}
\eta_Y^{ijk} = \frac{C(i)+C(j)+C(k)}{C(G)},~~~
\eta_\mu^{ij} = \frac{C(i)+C(j)}{C(G)}~.
\label{eta}
\end{eqnarray}
Therefore, these parameters can become very large
if $g(M_{\rm PL})/g(M_{\rm GUT})$ is large.
A rough  estimate shows that
\begin{eqnarray}
\frac{g(M_{\rm GUT}}{g(M_{\rm PL})} &\simeq &
\left[ \frac{ C(G)X_{\delta} \alpha_{\rm GUT}}{\pi \delta}
\right]^{1/2}
\left(\frac{M_{\rm PL}}{M_{\rm GUT}}\right)^{\delta/2}
\simeq 3.5(32)~~\mbox{for} ~~
    \delta=1(2),
  \label{ratio-g}
\end{eqnarray}
where we have used 
$\alpha_{\rm GUT}=0.04,~M_{\rm PL}/M_{\rm GUT}=10^2,~
G=SU(5)$
to obtain
the concrete  numbers.
These numbers should be compared with $1.3$
in the corresponding four-dimensional case \cite{ross}.

In contrast to $g,~Y^{ijk},~\mu^{ij}, ~M$,
the SSB parameters $B^{ij},~h^{ijk}$ and $(m^2)^{i}_{j}$
have a completely different behavior. We find that
the ratios of the SSB  parameters to the gaugino mass $M$
approach to their infrared attractive fixed points:
\begin{eqnarray}
B^{ij} / M  \mu^{ij} &\to & -\eta_\mu^{ij} ~, ~
h^{ijk} / M  Y^{ijk}  \to  -\eta_Y^{ijk}~, ~
(m^2)^{i}_{j}/ |M|^2 \to  \frac{C(i)}{C(G)}\delta_j^i, 
\label{ssb4}
\end{eqnarray}
where $\eta$'s are defined in (\ref{eta}).
Note that so far no assumption on the reality of the
SSB parameters has been made, 
and we recall that 
the phase of $M$ and $\mu^{ij}$  can always
be rotated away by a phase rotation that correspond 
to the $R$-symmetry  and 
an appropriate rotation  of the chiral superfields $\Phi$,
respectively. So, after these rotations,
 all the phases of $M$ and $\mu^{ij}$  are
transferred to those of $Y^{ijk}, h^{ijk}, B^{ij}$ and
$(m^2)^{i}_{j}$. 
Therefore, we may assume without
loss of generality that 
$M$ and $\mu^{ij}$ are real.
We see from (\ref{ssb4}) that
the low-energy structure is
completely fixed by the group theoretic structure of the model.
Furthermore, since $h^{ijk}$ and $(m^2)^{i}_{j}$
become aligned in the infrared limit,
{\it i.e.}, $h^{ijk} \propto Y^{ijk}$ and $(m^2)^{i}_{j} \propto
\delta_j^i$, the infrared forms (\ref{ssb4})
give desired initial values of the parameters at $M_{\rm GUT}$
to suppress FCNC processes  in the MSSM, and they predict that the
only CP-violating phase is the 
usual CKM phase \footnote{Eq. (\ref{ssb4}) means 
that the phases of $(h/MY)$ and $(B/M \mu)$
that cannot be rotated away approach zero
in the exact infrared limit.}.

One can easily estimate how much of
a disorder in the initial values at $M_{\rm PL}$
can survive at $M_{\rm GUT}$.
Suppose that there exists an $O(1)$ disorder in $(m^2)^i_j/|M|^2$.
Using the $\beta$ functions
(\ref{bg1}) and (\ref{bm21}), we find the deviation from
(\ref{ssb4}) to be
\begin{equation}
\left(\frac{g(M_{\rm PL} )}{g(M_{\rm GUT} )}\right)^4
\left[\frac{(m^2)^i_j}{|M|^2}(M_{\rm PL})
-\frac{C(i)}{C(G)}\delta^i_j\right].
\label{disorderm}
\end{equation}
Then inserting the value of $ g(M_{\rm PL} )/g(M_{\rm GUT})$
given in (\ref{ratio-g}),
we find that 
an $O(1)$ disorder at $M_{\rm PL}$ becomes
a disorder of $O(10^{-2})$ and $O(10^{-6})$  at $M_{\rm GUT}$
for $\delta=1$ and $2$,
respectively. Note that the off-diagonal elements of $(m^2)^i_j$
as well as the differences among the diagonal elements
$\Delta m^2(i,j)=(m^2)^i_i-(m^2)^j_j$ (if $C(i)=C(j)$)
belong to the disorder. However, their contributions to
$(\delta_{ij})_{LL,RR}$ of \cite{fcnc} are less than
$O(10^{-6})$ for $\delta=2$,  and therefore
the most stringent constraints coming from
the $K_S-K_L$ mass difference $\Delta m_K$ and the decay
$\mu\to e\gamma$ are
satisfied \cite{fcnc}.
In the case of five dimensions ($\delta=1$) the
suppression of the disorder will be sufficient,
if the gauginos are much heavier than
the sfermions \cite{fcnc}.
[If we use $M_{\rm PL}/M_{\rm GUT} \sim 10^3$, then the suppression
is much improved.]

Similarly, using (\ref{bg1}) and (\ref{bh1}), we obtain
the deviation for the tri-linear couplings from (\ref{ssb4}) as
\begin{eqnarray}
\left(\frac{g(M_{\rm PL} )}{g(M_{\rm GUT} )}\right)^2
\left[\frac{h^{ijk}}{M Y^{ijk}}(M_{\rm PL})
+\eta^{ijk}_Y  (M_{\rm PL})\right],
\label{disorderh}
\end{eqnarray}
where use has been made of (\ref{yijk}).
Suppose the tri-linear couplings to be order of $M Y^{ijk}$
at $M_{\rm PL}$.
Then we find that 
\begin{equation}
\left|
\frac{h^{ijk}}{M Y^{ijk}}(M_{\rm GUT}) + \eta_Y^{ijk}
\right|
\lsim 
\left(\frac{g(M_{\rm PL})}{g(M_{\rm GUT})}\right)^{2+\eta_Y^{ijk}}.
\label{disorderh2}
\end{equation}
Note that the phases of $h^{ijk}/M Y^{ijk}$ are also suppressed.
In the case of $G=SU(5)$, $\eta_Y^{ijk}=48/25 (42/25)$
for the up (down) type Yukawa couplings.
Using  (\ref{ratio-g}) again, we find that the right-hand side of
(\ref{disorderh2}) is
$\sim 10^{-2(6)}$ for $\delta=1 (2)$.
This disorder contributes, for instance,
to $Im (\delta_{ii})_{LR}$
as well as $Re(\delta_{ij})_{LR}$ of \cite{fcnc}.
Therefore  our  suppression  mechanism  can satisfy
the most stringent constraints coming from
the electric dipole moments (EDM) of the neutron and the
electron and also from $\epsilon'/\epsilon$ in the
$K^0-\bar{K^0}$ mixing \cite{fcnc}.
Similarly the phases of the B-parameters, $B^{ij}/M \mu^{ij}$
are also suppressed.
\begin{figure}
\epsfxsize=11cm
\centerline{\epsfbox{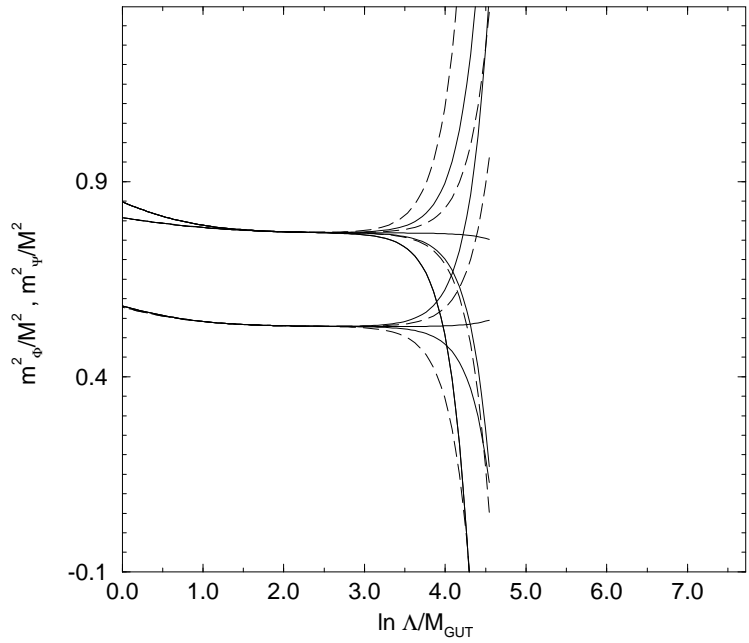}}
\caption{Infrared attractiveness of  $m^2_{5}/|M|^2$ and
$m^2_{10}/|M|^2$. The dashed (solid)  lines correspond to
the third (first two) generation(s).
$m^2_{10^{1,2}}/|M|^2 > m^2_{10^{3}}/|M|^2
> m^2_{5^{1,2}}/|M|^2 \simeq m^2_{5^{3}}/|M|^2$
at $\Lambda=M_{\rm GUT}$.}
\label{fig1}
\end{figure}

In concrete examples, there will be logarithmic corrections
to (\ref{ssb4}) to which the Yukawa couplings $Y^{ijk}$
non-trivially contribute.
How much 
the logarithmic corrections can
amplify the disorder
will  be model-dependent.
In \cite{kubo} we performed detailed analyses on the
logarithmic corrections in a GUT model based on $SU(5)$.
Fig.~1, which is one of the main results of \cite{kubo},  shows the evolution
of $m_{5}^{2}/|M|^2$ and
$m_{10}^{2}/|M|^2$, respectively,
in an obvious notation.
The dashed lines correspond to
the third generation.
The differences
$\Delta m^2_{10} (i,3)/|M|^2=
|m_{10^{i}}^{2}-m_{10^{3}}^{2}|/|M|^2$
with $i=1,2$ directly 
contribute to, for instance,  $ \Delta m_B$ in the $B-\bar{B}$ mixing as well as  to 
$\tau\to e\gamma$ and $\tau\to \mu\gamma$. We find that
$\Delta m^2_{10} (i,3)/|M|^2 \lsim 0.04$ at $M_{\rm GUT}$,  
which means that 
$|(\delta_{13,23}^{\l,u})_{RR}|,
|(\delta_{13,23}^{d,u})_{LL}|
\lsim \times 10^{-2}$ at $M_{\rm GUT}$.
Therefore,  
$ \Delta m_B$  and 
$\tau\to e\gamma$ and $\tau\to \mu\gamma$ are sufficiently suppressed.
The differences
$\Delta m^2_{10} (i,3)/|M|^2$ also contribute through the 
the mixing between the first two generations and the third generation 
to $\Delta m_K$ and $\mu\to e\gamma$. 
Assuming that the mass matrix of the up-type quarks is diagonal,
and using the known values of
Cabibbo-Kobayashi-Maskawa matrix $V_{CKM}$,
we find that that
$\Delta m^2_{\Psi} (i,3)/|M|^2 \lsim 0.04$  does not
cause any problems with the FCNC processes mentioned above.
The difference of $-0.04$ in $m_{\Psi}^{2}/|M|^2$ also causes
no problem for $b \to s \gamma$ \cite{fcnc}.
We also studied the logarithmic contributions
to  the non-aligned part of $h^{ijk}$,
which may contribute to
the EDMs as well as $\epsilon'/\epsilon$ in the $K^0-\bar{K^0}$ 
system \cite{fcnc}. It is found that 
the disorder of the trilinear couplings
caused by the  Yukawa couplings
are sufficiently suppressed
to satisfy the constraints coming from these parameters.

We conclude that gauge interactions in extra dimensions
can be used to suppress the disorder of the SSB terms at the 
fundamental scale so that the FCNC processes and dangerous
CP-violating phases become tiny at lower energy scales.
Moreover, no charged sparticles become tachyonic
in this scenario of the SSB parameters.
The suppression mechanism of the FCNC and CP-phases presented 
here
does not properly work in four dimensions. Therefore, the smallness of FCNC
as well as of EDM is a possible hint of the  existence of extra dimensions.

\vspace{0.5cm}
This work is supported by the Grants-in-Aid
for Scientific Research  from
 the Japan Society for the Promotion of Science (JSPS) (No. 11640266,
 No. 13135210).
We would like to thank T.~Ishikawa,
Y.~Kajiyama, T.~Kobayashi, M.~Mondragon and H.~Nakano.
for useful discussions.

\end{document}